\documentclass{moriond}
\def\Journal#1#2#3#4{{#1} {\bf #2}, #3 (#4)}

\def\PRD{{\em Phys. Rev.} D}

\def\be{\begin{equation}}
\def\ee{\end{equation}}
\def\bea{\begin{eqnarray}}
\def\eea{\end{eqnarray}}

\def\symff{{\rm SYM}_{4,4}}
\def\symod{{\rm SYM}_{1,{\cal D}}}
\def\symot{{\rm SYM}_{1,10}}
\newcommand{\Photo}{}

\begin{document}
\vspace*{4cm}
\title{Effective field theory treatment of ${\cal N}=4$ supersymmetric Yang-Mills thermodynamics}
\author{Ubaid Tantary\,\footnote{Presenter}, Jens O. Andersen, Qianqian Du, and Michael Strickland }
\address{Department of Physics, Kent State University,\\
800 E Summit St, Kent, USA\\ \ }
\maketitle
\abstracts{At finite temperature the free energy density of ${\cal N}=4$ supersymmetric Yang-Mills can be calculated using resummed perturbation theory through the order $\lambda^{5/2}$.  Effective field theory methods provide a useful alternative approach to streamline these calculations. In this proceedings contribution, I review recent work with my collaborators where we used effective field theory methods to calculate the free energy density of ${\cal N}=4$ supersymmetric Yang-Mills in four spacetime dimensions through second order in the 't Hooft coupling $\lambda$. At this order the contributions to the free energy density come from the hard scale $T$ and the soft scale $\sqrt{\lambda}T$. The contribution from the scale $T$ enters through the coefficients in the effective Lagrangian obtained by dimensional reduction and the effects of the scale $gT$ can be calculated using perturbative methods in the effective theory.}

\section{INTRODUCTION}

One of the motivations to study the thermal properties of ${\cal N}=4$ supersymmetric Yang-Mills in four dimensions ($\symff$) is that, at finite temperature, the weak-coupling limit of this theory has many similarities with  quantum chromodynamics (QCD). At large-$N_c$ one can make use of the AdS/CFT correspondence to obtain the strong coupling result.\cite{KRT} In a recent paper my collaborators and I were able to compute the free energy density in the opposite, weak-coupling regime through ${\cal O}(\lambda^2)$ using resummed perturbation theory using techniques developed by Arnold and Zhai in quantum chromodynamics (QCD).\cite{ma,AJ} In a followup work we used effective field theory (EFT) methods to reproduce the perturbative expansion of the free energy through the same order.\cite{jo}  The advantage of using EFT methods is that they can be more straightforwardly extended to higher orders in the gauge coupling. We based our EFT calculations on the methods developed by Braaten and Nieto to calculate the thermodynamics of QCD through ${\cal O}(\lambda^{5/2})$.\cite{bd}

\section{SUPERSYMMETRIC YANG-MILLS THEORY}
The $\symff$ theory can be obtained by dimensional reduction of $\symod$ in \mbox{$\mathcal{D}=\mathcal{D}_{\textrm{max}}=10$} with all fields being in the adjoint representation of $SU(N_c)$. The Lagrangian that generates the perturbative expansion for $\symff$ in Minkowski-space can be expressed as
 \bea
  \mathcal{L}_{\symff}&=&\textrm{Tr}\bigg[{-}\frac{1}{2}G_{\mu\nu}^2+(D_\mu\Phi_A)^2+i\bar{\psi}_i { D}\psi_i-\frac{1}{2}g^2(i[\Phi_A,\Phi_B])^2 \nonumber \\
  &&  - i g \bar{\psi}_i\big[\alpha_{ij}^{\texttt{p}} X_{\texttt{p}}+i \beta_{ij}^{\texttt{q}}\gamma_5Y_{\texttt{q}},\psi_j\big] \bigg] +\mathcal{L}_{\textrm{gf}}+\mathcal{L}_{\textrm{gh}}+\Delta\mathcal{L}_{\textrm{SYM}} \, ,
 \eea
with $\Phi_A \in (X_1,Y_1,X_2,Y_2,X_3,Y_3)$,  where $X$ and $Y$ denote scalar and pseudoscalar fields, respectively, and all fields are in the adjoint representation.\cite{ma,jo}

\section{DIMENSIONAL REDUCTION AND EFT TECHNIQUE}
The calculation of $\symff$ thermodynamics requires two types of dimensional reduction:  (1) the equivalence between ten-dimensional $\symot$ and four-dimensional $\symff$ upon dimensional reduction, and (2) the additional dimensional reduction of $\symff$ to three dimensions that occurs at high temperatures. The latter is based on the old idea that static properties in (3+1)D field theory can be expressed in terms of an EFT in three spatial dimensions written in terms of the bosonic zero modes. 

The construction of the SUSY-EFT involves writing down the partition function in the full theory and then integrating out non-static modes to obtain the partition function in (electric) SUSY-EFT
\bea
{\cal Z}_{\symff} &=& \! \int \!
{\cal D}\bar{\eta}{\cal D}\eta
{\cal D}\bar{\psi}_i{\cal D}\psi_i
{\cal D}A_{\mu}^a{\cal D}\Phi_A^a \, e^{- \!{\int_0^{\beta}} d\tau\int d^3x \, {\cal L}_{\symff}} \, , \\
{\cal Z}_{ESYM} &=& \int{\cal D}\bar{\eta}{\cal D}{\eta}{\cal D}A_0^a{\cal D}A_i^a{\cal D}\Phi_A^ae^{-f_EV-\int \!d^3x\,{\cal L}_{\rm ESYM}} \,,
\eea
where the partition functions contain fields in full and effective theory, respectively, and $f_E$ is the coefficient of the unit operator.
${\cal L}_{\rm ESYM}$ is given by the most general Lagrangian that can be constructed from the
fields $A_i^a$, $A_0^a$, and $\Phi_A^a$.  Through the order in the t'Hooft coupling required one has
\bea\nonumber
{\cal L}_{\rm ESYM}&=&
\frac{1}{2}{\rm Tr}\left[G_{ij}^2\right]
+{\rm Tr}[(D_iA_0)(D_iA_0)]
+{\rm Tr}[(D_i\Phi_A)(D_i\Phi_A)]
+m_E^2{\rm Tr}[A_0^2]
+m_S^2{\rm Tr}[\Phi_A^2]
\nonumber \\ &&
+h_E{\rm Tr}[(i[A_0,\Phi_A])^2]
 +\frac{1}{2}g^2_3{\rm Tr}[(i[\Phi_A,\Phi_B])^2]
+{\cal L}_{\rm gf}+{\cal L}_{\rm gh}+
\delta{\cal L}_{\rm ESYM} \, , 
\label{lagesym}
\eea
where $A_0=t^aA_0^a$, $\Phi_A=t^a\Phi_A^a$, $(D_iA_0)^a=\partial_iA_0^a+g_E f^{abc}A_i^b A_0^c$, and
$G_{ij}^a=\partial_i A_j^a-\partial_j A_i^a+g_Ef^{abc}A_i^bA_j^c$ is the nonabelian
field strength with gauge coupling $g_E$ and $f^{abc}$ are the structure constants. 
\section{PARAMETERS OF THE EFFECTIVE
THEORY}
In this section, I briefly outline the procedure to determine the parameters of the effective theory to the order $\lambda$ needed to calculate the free energy to order $\lambda^{2}$.

\subsection{Coefficient of unit operator}

$f_E$ is the coefficient of the unit operator and can be interpreted as the contribution to the free energy from the hard scale $T$. The hard or unresummed contributions are calculated in the full $\symff$ theory. For the unresummed (hard) contributions we do not need to consider the thermal masses of the gluons, fermions, or scalars.  As a result, we can calculate all the hard contributions using SUSY dimensional reduction from ${\rm \symot}$ to $\symff$, requiring dramatically fewer Feynman diagrams. I will only list the contribution from the 3-loop diagrams because they possess an uncancelled divergence and outline its systematic treatment. 
\begin{figure}[h!]
  \centering
  \includegraphics[width=0.6\textwidth]{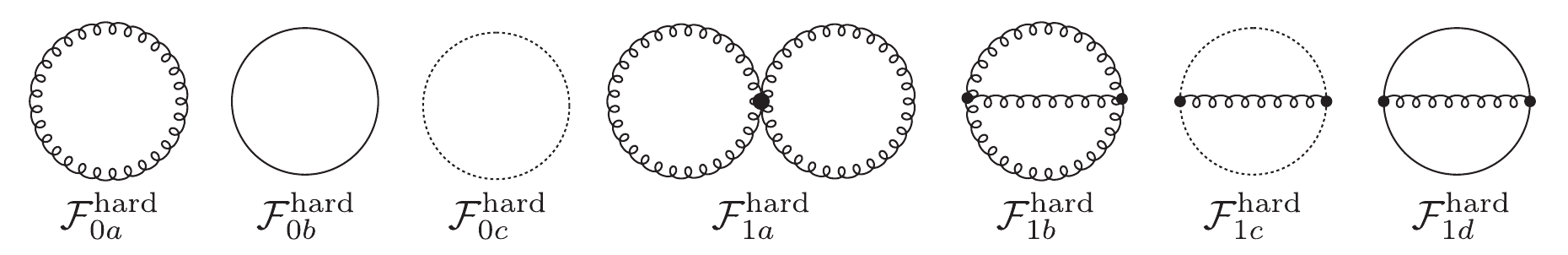}
  \includegraphics[width=0.82\textwidth]{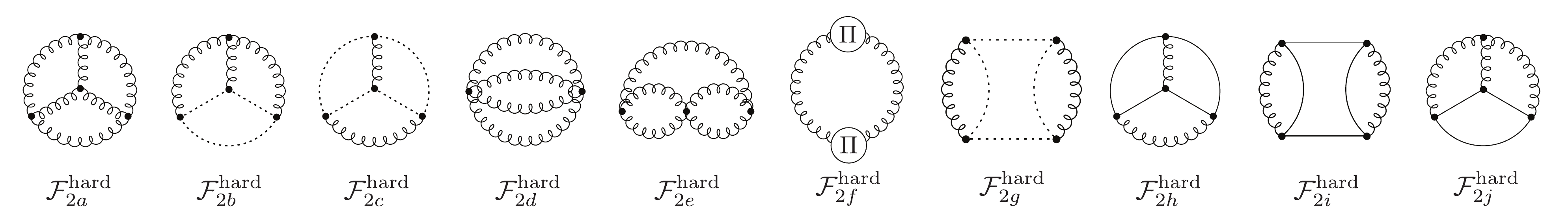}
\vspace{-2mm}
\caption{One-, two-loop, and three-loop diagrams contributing to the ${\rm \symot}$ free energy density. }  \end{figure}
\be
{\cal F}_2^{\rm hard}=
-d_A\left({\pi^2T^4\over6}\right)\left[
{3\over4\epsilon}+{9\over2}\log{\Lambda\over4\pi T}
\right. \nonumber \\  \label{f33} \left.
+{3\over2}\gamma_E
+3{\zeta^{\prime}(-1)\over\zeta(-1)}
+{15\over4}-\log2
\right]\left({\lambda\over\pi^2}\right)^2 \!. \;\;\;\;
\ee

The remaining pole is cancelled by the counterterm $\delta f_E$ for the coefficient of the unit operator $f_E$. At the order required, the counterterm is a polynomial in $m_E$, $m_S$, $g_E$, $h_E$, and $g_3$.  We find
 \be
 \delta f_E=-{d_AN_c\over4(4\pi)^2\epsilon}g_E^2\left[m_E^2+6m_S^2\right]. 
 \ee

The final result ($
{\cal F}_0^{\rm hard}+{\cal F}_1^{\rm hard}+
{\cal F}_2^{\rm hard}-T\delta f_E$) for the renormalized unit operator $f_E$ is  
\bea
f_E(\Lambda)T\label{renhard}=
-d_A{\pi^2T^4\over6}\left\{
1-{3\over2}{\lambda\over\pi^2}+\left[3\log{\Lambda\over4\pi T}
+{39\over16}+{3\over2}\gamma_E
+{3\over2}{\zeta^{\prime}(-1)\over\zeta(-1)}-{1\over2}\log2
\right]\left({\lambda\over\pi^2}\right)^2\right\} .
\eea

\subsection{Mass parameters}

We calculate the coefficient $m_E^2$ and $m_S^2$   of the terms $A_0^a A_0^a$ and $\Phi_A^a \Phi_A^a$  in the ESYM lagrangian to one-loop order. Their physical interpretation is that they give the contribution to the static screening masses
from the hard scale $T$.

\subsection{Coupling constants} 
Simply substitute $A_0^a({\bf x},\tau)\rightarrow \sqrt{T}A_0^a({\bf x})$ and $\Phi_A^a({\bf x},\tau)\rightarrow\sqrt{T}\Phi_A^a({\bf x})$ in the full theory and compare $\int_0^{\beta} d\tau {\cal L}_{\rm SUSY}$ with
the effective theory, ${\cal L}_{\rm ESYM}$ to yield: $g_E^2=g^2T$, $g^2_3=g^2T$    and $h_E=g^2T.$

\section{CALCULATIONS IN THE EFFECTIVE FIELD THEORY}

In the EFT technique the soft contribution is obtained by performing two-loop perturbative calculations in the effective theory. Denoting the contribution from 
the soft scale $\sqrt{\lambda}T$ by $f_M$, we have $f_M=-{\log{\cal Z}_{\rm ESYM}\over V}$.
\begin{figure}[b]
  \centering
  \includegraphics[width=0.35\textwidth]{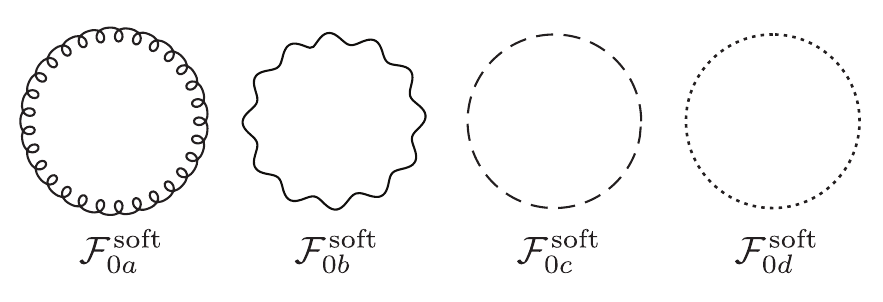}
   \includegraphics[width=0.4\textwidth]{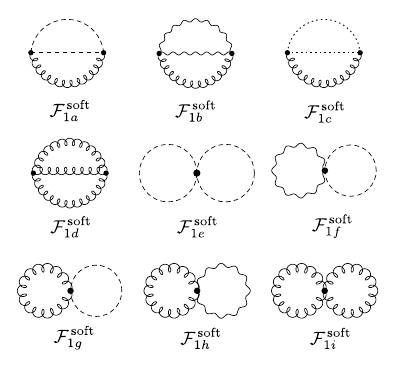}
  \caption{The one- and two-loop soft  contributions to the $\symff$ free enrgy. }
\end{figure}
There is an ultraviolet divergence in the two-loop contribution which  requires renormalization, cf. renormalization of $f_E$.
The divergence is cancelled by the counterterm $
\delta f_E = -{d_AN_c\over4(4\pi)^2\epsilon}g_E^2\left[m_E^2
+6m_S^2\right]$ and the total soft contribution is found to be
\be
f_M=
\label{rensoft}
-d_A{\pi^2T^3\over6}\left\{
(3+\sqrt{2})\left({\lambda\over\pi^2}\right)^{3\over2}
+ \left[
-3\log{\Lambda\over4\pi T}-{81\over16}-{9\sqrt{2}\over8}-{21\over8}\log2+{3\over2}\log{\lambda\over\pi^2}
\right]\left({\lambda\over\pi^2}\right)^{2}
\right\} \! .
\ee
 Adding Eqs.~(\ref{renhard}) and~(\ref{rensoft}), we obtain our final result\,\cite{jo}
\bea
{\cal F}_{0+1+2}&=&(f_E+f_M)T=-d_A{\pi^2T^4\over6}\left\{
1-{3\over2}{\lambda\over\pi^2}+(3+\sqrt{2})\left({\lambda\over\pi^2}\right)^{3\over2}
\right.  \nonumber \\  && \hspace{1cm} \left.
  +\left[-{21\over8}-{9\sqrt{2}\over8}+{3\over2}\gamma_E
+{3\over2}{\zeta^{\prime}(-1)\over\zeta(-1)}
-{25\over8}\log2+{3\over2}
\log{\lambda\over\pi^2}
\right]\left({\lambda\over\pi^2}\right)^{2}
\right\}\;.
\label{dgh}
\eea
 This is the complete result for the free energy through order $\lambda^2$ for general $N_c$.
It is in agreement with the result found earlier using resummed perturbation techniques.\,\cite{ma}

\section{Conclusion and Oulook}

\begin{figure}[t!]
  \centering
  \includegraphics[width=0.525\textwidth]{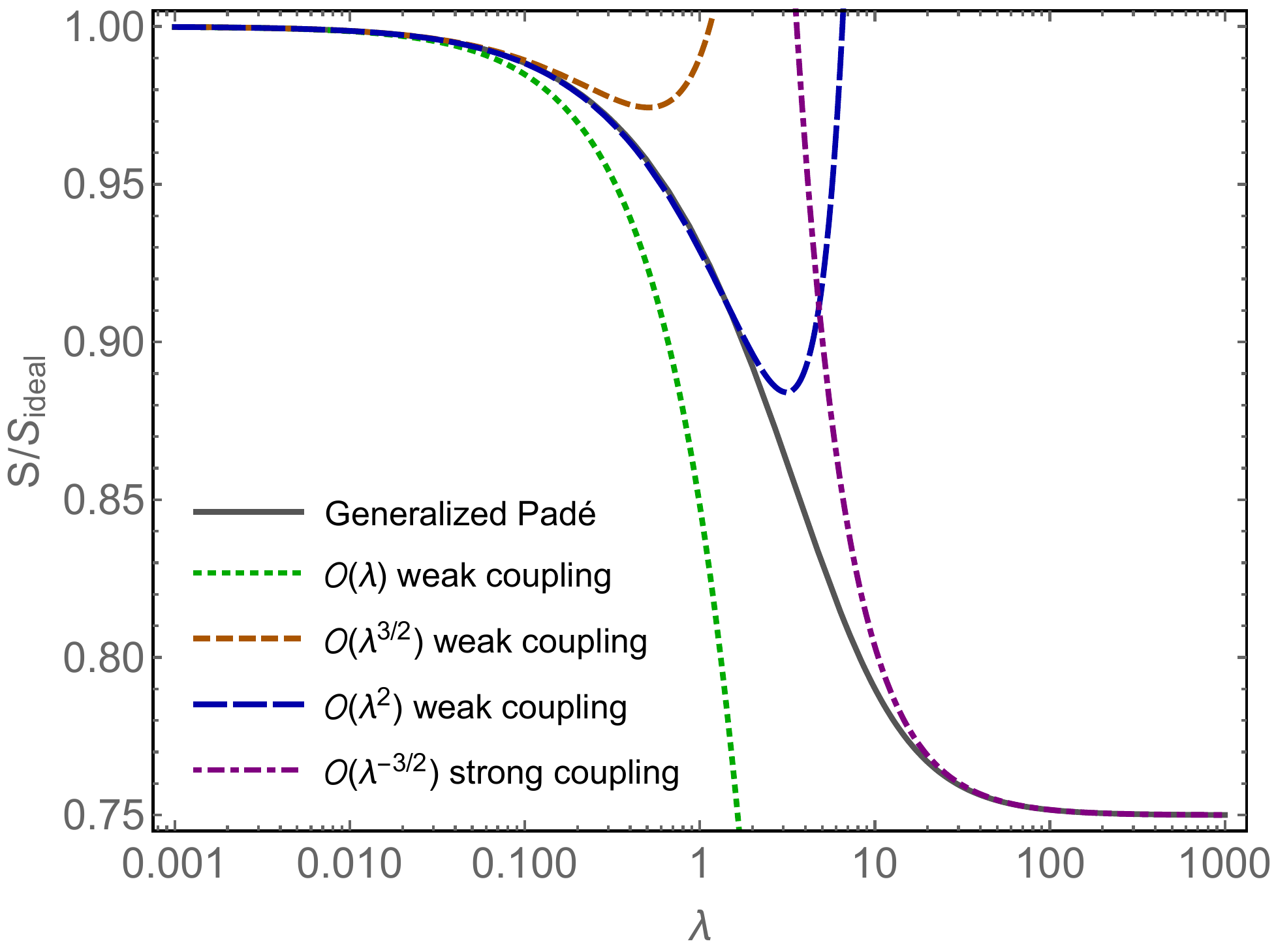}
  \caption{The entropy density ${\cal S}$ normalized by the ${\cal S}_{\rm ideal}$
in ${\rm \symff}$ as a function of the t 'Hooft coupling $\lambda$. }
\label{fig:edens}
\end{figure}

In this work, I reviewed the computation of the $\symff$ thermodynamic functions to ${\cal O}(\lambda^2)$ using EFT techniques. The final result, presented in Eq.~(\ref{dgh}), confirms our previous result\,\cite{ma} and extends our knowledge of weak-coupling $\symff$ thermodynamics to include terms at ${\cal O}(\lambda^2)$  and ${\cal O}(\lambda^2 \log\lambda)$.  With the ${\cal O}(\lambda^2)$  and ${\cal O}(\lambda^2 \log\lambda)$ coefficients in the $\symff$ free energy, we then constructed a large-$N_c$ Pad\'{e} approximant that interpolates between the weak- and strong-coupling limits. Fig.~\ref{fig:edens} summarizes our findings.  The next term in the weak-coupling expansion will be of order $\lambda^{5\over2}$ and is the highest order that can be obtained using purely perturbative calculations. Computation of this term is a work in progress.

\section*{Acknowledgments}
 M.S. and U.T. were supported by the U.S. Department of Energy, Office of Science, Office of Nuclear Physics under Award No.~DE-SC0013470.

\end{document}